\documentclass{article}

\usepackage{graphicx}
\usepackage{booktabs} 

\usepackage{hyperref}
\usepackage{booktabs} 
\usepackage{multirow} 
\usepackage{siunitx}  
\usepackage{graphicx}   
\usepackage{subcaption} 
\usepackage{float}
\usepackage{tabularx}

\usepackage[preprint]{icml2026}

\usepackage{amsmath}
\usepackage{amssymb}
\usepackage{mathtools}
\usepackage{amsthm}
\usepackage{longtable}
\usepackage{booktabs}

\usepackage[capitalize,noabbrev]{cleveref}

\usepackage{graphicx}
\usepackage{kotex}
\usepackage{booktabs}                                       
\usepackage{multirow}                                       
\usepackage{array}                                          
\usepackage{fontawesome}                                    
\usepackage[table]{xcolor}                                  
\usepackage{colortbl}                                       
\usepackage{hyperref}

\usepackage[most]{tcolorbox}
\tcbuselibrary{breakable,skins,listingsutf8}
\usepackage{listings}

\lstdefinestyle{promptlisting}{
  basicstyle=\ttfamily\footnotesize,
  columns=fullflexible,
  keepspaces=true,
  showstringspaces=false,
  tabsize=2,
  breaklines=true,
  breakatwhitespace=false
}

\newtcblisting{promptbox}[2][]{
  enhanced,
  breakable,
  width=\linewidth,
  colback=gray!10,
  colframe=black,
  boxrule=0.5mm,
  sharp corners,
  left=1.5mm,right=1.5mm,top=1mm,bottom=1mm,
  colbacktitle=black,
  coltitle=white,
  fonttitle=\ttfamily\small,
  title={\parbox{\dimexpr\linewidth-3mm\relax}{\raggedright #2}}, 
  listing only,
  listing options={style=promptlisting},
  #1
}

\definecolor{linkpink}{HTML}{E91E63} 

\newtcbox{\codebadge}{
  on line,
  arc=2.2pt,
  colback=linkpink!8,
  colframe=linkpink,
  colupper=linkpink,
  boxrule=0.4pt,
  left=6pt,right=6pt,top=2pt,bottom=2pt
}

\newcommand{\squishlist}{
   \begin{list}{$\bullet$}
    { \setlength{\itemsep}{1pt}
      \setlength{\parsep}{0pt}
      \setlength{\topsep}{2pt}
      \setlength{\partopsep}{0pt}
      \setlength{\listparindent}{-2pt}
      \setlength{\itemindent}{-5pt}
      \setlength{\leftmargin}{1.5em}
      \setlength{\labelwidth}{0em}
      \setlength{\labelsep}{0.5em}
    }
}
\newcommand{\squishlistend}{
    \end{list}  }
\newcommand{\squishend}{
    \end{list}  }


\theoremstyle{plain}

\theoremstyle{definition}

\theoremstyle{remark}

\usepackage{capt-of} 

\usepackage[textsize=tiny]{todonotes}
\usepackage{dsfont}
\usepackage{xspace}

\newcommand{\Ours}{FiF\xspace}
\newcommand{\OurFullName}{Failure is Feedback\xspace}
\newcommand{\Title}{History-Aware Backtracking for Agentic Traversal in Multimodal Graphs\xspace}

\newcommand{\MultimodalQA}{\textsc{MultimodalQA}$^{\texttt{Doc}}$\xspace}
\newcommand{\MMCoQA}{\textsc{MMCoQA}$^{\texttt{Doc}}$\xspace}
\newcommand{\WebQA}{\textsc{WebQA}$^{\texttt{Doc}}$\xspace}

\icmltitlerunning{\textsc{\OurFullName}: \Title}
\icmlsetsymbol{advising}{$\dagger$}

\newcommand{\icmlEqualAdvising}{\textsuperscript{$\dagger$} Advising}

\begin{document}

\twocolumn[{
\icmltitle{
\texorpdfstring
{\textsc{\OurFullName}: \Title}
{\textsc{\OurFullName}: \Title}
}

\icmlsetsymbol{equal}{*}
\icmlsetsymbol{advising}{, $\dagger$}

\begin{icmlauthorlist}

\icmlauthor{Joohyung Yun}{postech_cse}
\icmlauthor{Doyup Lee}{directorlabs}
\icmlauthor{Wook-Shin Han}{postech_gsai,advising}

\end{icmlauthorlist}

\icmlaffiliation{postech_gsai}{GSAI, POSTECH}
\icmlaffiliation{postech_cse}{CSE, POSTECH}
\icmlaffiliation{directorlabs}{DirectorLabs}
\icmlcorrespondingauthor{Joohyung Yun}{jhyun@dblab.postech.ac.kr}
\icmlkeywords{Machine Learning, ICML}



\vskip 0.3in
}]




\begin{NoHyper}
\printAffiliationsAndNotice{\icmlEqualAdvising}
\end{NoHyper}

\begin{abstract}

Open-domain multimodal document retrieval aims to retrieve specific components (paragraphs, tables, or images) from large and interconnected document corpora.
Existing graph-based retrieval approaches typically rely on a uniform similarity metric that overlooks hop-specific semantics, and their rigid pre-defined plans hinder dynamic error correction.
These limitations suggest that a retriever should adapt its reasoning to the evolving context and recover intelligently from dead ends.
To address these needs, we propose \textsc{Failure is Feedback} (\textsc{FiF}), which casts subgraph retrieval as a \textit{sequential decision process} and introduces two key innovations.
(i) We introduce a \textit{history-aware backtracking mechanism}; unlike standard backtracking that simply reverts the state, our approach piggybacks on the context of failed traversals, leveraging insights from previous failures. 
(ii) We implement an \textit{economically-rational agentic workflow}. 
Unlike conventional agents with static strategies, our orchestrator employs a cost-aware traversal method to dynamically manage the trade-off between retrieval accuracy and inference costs, escalating to intensive LLM-based reasoning only when the prior failure justifies the additional computational investment.
Extensive experiments show that \textsc{\Ours} achieves state-of-the-art retrieval on the benchmarks of \textsc{MultimodalQA}, \textsc{MMCoQA} and \textsc{WebQA}.

\end{abstract}
\vspace{-2em}

\section{Introduction}
\label{sec:introduction}
\vspace{-0.5em}

\begin{figure}[t]
  \centering
  \includegraphics[width=\linewidth]{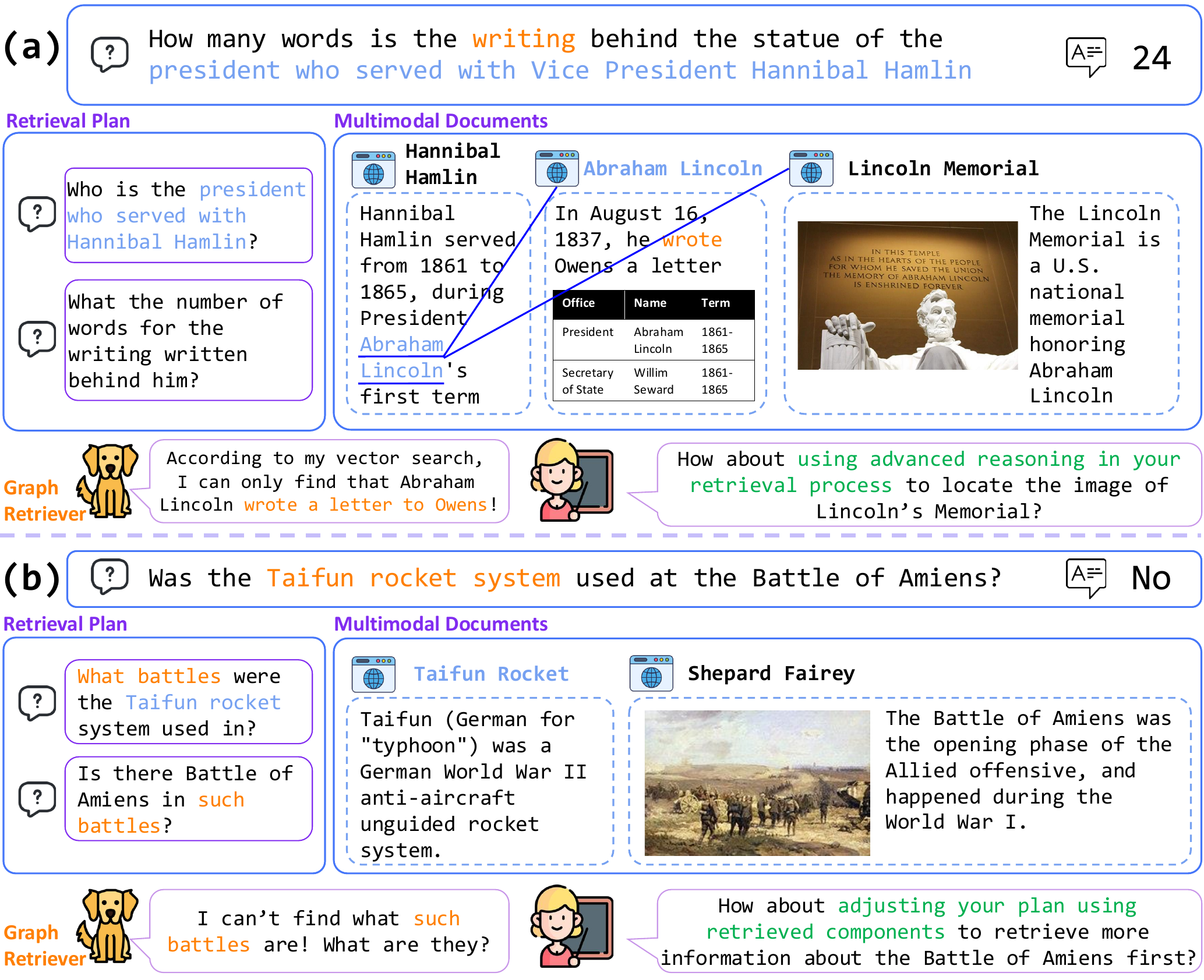}
  \vspace{-4mm}
  \caption{
    Motivating examples of multihop retrieval failures in existing graph retrieval approaches.
    (a) Vector-similarity-driven traversal follows a spurious cue.
    (b) Fixed retrieval plan produces an underspecified hop and fails to recover from a dead end.
  }
  \label{fig:motivating_example_figure}
  \vspace{-6mm}
\end{figure}

Searching the web has become a part of everyday life.
This routine increasingly underpins multimodal retrieval-augmented generation (RAG), where a model answers a user query by grounding its output in retrieved evidence~\cite{4_colpali, awakening}.
In practice, much of this evidence lives in webpages or PDFs---multimodal documents with three salient characteristics:
(i) each document is composed of multimodal \textit{components} (paragraphs, tables, and images);
(ii) the meaning of a component is often shaped by local document context (e.g., captions and surrounding components); and 
(iii) components are connected through explicit signals (hyperlinks, cross-references) as well as implicit signals (e.g., same-section adjacency).
Moreover, documents themselves are linked via hyperlinks and citations, forming a large graph that users implicitly navigate while browsing.
We refer to the resulting setting as \emph{open-domain multimodal document retrieval} (OMDR): given a query, the system must return a small ranked set of relevant components from this large, noisy, and interlinked graph, often requiring multihop and multimodal exploration~\cite{1_lilac, 9_ircot, 4_colpali}.

Given these intricate characteristics of OMDR, representing a document collection as a graph has emerged as a powerful paradigm for capturing the multi-granularity and interconnectedness of multimodal evidence~\cite{1_lilac}.
It shows the pros of preserving the structural dependencies and navigational scaffolds inherent in webpages, which is required when navigating heterogeneous components.
The most recent work introduces the \textit{layered component graph}, which organizes components together with their constituent subcomponents (sentences, table rows, and image objects)~\cite{1_lilac}.
In this formulation, \emph{navigational edges} encode relations among components (e.g., hyperlinks, same-section adjacency), while \emph{hierarchical edges} connect each component to its subcomponents.
By jointly modeling these edge types across layers, a retriever can traverse component-to-component paths for multihop exploration and move up/down the hierarchy to operate at the appropriate granularity.

While graph-based structures provide a rich representation of multimodal evidence, existing retrieval algorithms often struggle to fully exploit this potential due to operational rigidity.
In particular, (a) traversal is typically driven by a single, hop-agnostic embedding-based scoring rule and (b) executed with a largely pre-specified procedure, limiting dynamic error correction.
Figure~\ref{fig:motivating_example_figure} highlights these failures.
In Figure~\ref{fig:motivating_example_figure}(a), it follows a superficially related textual cue and retrieves an irrelevant snippet, failing to ground on the crucial visual evidence.
In Figure~\ref{fig:motivating_example_figure}(b), it issues an underspecified follow-up (``such battles'') and gets stuck in a dead end, rather than adapting its trajectory after the failure.
As a result, once the retriever follows a spurious edge or reaches a dead end, errors propagate across hops and degrade final retrieval quality~\cite{9_ircot, 10_selfrag}.
Moreover, these methods lack a principled mechanism for deciding \emph{when} expensive reasoning is warranted, leading to under-reasoning on ambiguous hops or over-spending computation across a trajectory.

To overcome these limitations, we argue that a retriever must evolve from a static path-follower into an adaptive decision-maker that navigates the graph through a sequential reasoning process.
Concretely, OMDR is naturally stateful: as evidence accumulates, the information need shifts, and failures reveal which interpretations, routes, or strategies are unproductive.
This suggests casting traversal as a \emph{sequential decision process} over an evolving information state, where each step chooses (i) what to ask next (subquery), (ii) how to retrieve (tool/strategy), and (iii) where to move (edge type and granularity) conditioned on the current evidence.
Achieving this requires addressing three coupled challenges.
First, edge-following is not merely similarity matching; it often requires high-level reasoning to judge whether a candidate node will lead to the final answer under the current context.
Second, the retriever must adapt to evolving context by refining hypotheses and subqueries, and by recovering from dead ends using failure signals rather than adhering to a fixed, pre-defined plan.
Third, it must balance accuracy and efficiency: while LLM-based reasoning can improve retrieval precision, it introduces substantial overhead, so the system must decide economically when to escalate from lightweight matching to intensive reasoning.

To address these needs, we propose \textsc{\textbf{\OurFullName}} (\textsc{\Ours}).
We formalize OMDR as a finite-horizon \emph{information-state MDP}, where the state is a structured memory that records accumulated evidence together with the history of attempted subqueries, strategies, and explicit success/failure outcomes.
This formulation turns graph traversal into an \emph{economically-rational} agentic workflow: at each hop, an orchestrator dynamically decides \emph{what to ask}, \emph{how to retrieve}, and \emph{where to move} given the current information state, rather than executing a rigid traversal recipe.
To realize cost-sensitive control, \textsc{\Ours} maintains a portfolio of strategies across an accuracy--efficiency spectrum, starting from low-cost vector matching and escalating to higher-cost LLM reasoning only when a hop is ambiguous or an attempt fails.
Finally, to make multihop navigation resilient in noisy open-domain graphs, \textsc{\Ours} introduces \emph{history-aware backtracking}: unlike standard backtracking that simply reverts the state, our approach piggybacks on failure traces to re-anchor the search to a more promising prior context, revise subsequent subqueries, and avoid repeating previously failed routing patterns.

In summary, we make three primary contributions:
\vspace{-2mm}
\squishlist
    \item [1.] We formulate the OMDR problem as a sequential decision process with economic rationality. We redefine OMDR as an information-state MDP, operationalizing it through an LLM-enabled agentic workflow that treats retrieval strategy as a dynamic choice.
    \item [2.] We propose dynamic cost-aware strategy escalation.
    We introduce a novel mechanism that maintains a portfolio of strategies across an accuracy-efficiency spectrum. 
    Our orchestrator avoids over-reasoning by starting with low-cost vector matching and only escalating to high-cost LLM reasoning when a hop is identified as ambiguous or follows a recorded failure.
    \item [3.] We propose history-aware backtracking for resilient navigation, which converts failed traversals into constructive feedback. 
    By piggybacking on failure traces, the orchestrator re-anchors its search to prior contexts while revising its subqueries and escalating its strategy, enhancing both robustness and efficiency.
\squishend
\vspace{-2mm}

\vspace{-1mm}
\section{Related Work}
\label{sec:relatedwork}
\vspace{-1mm}

\vspace{-1mm}
\subsection{Multimodal Retrieval Methods}
\label{sec:related_multimodal_retrieval}
\vspace{-1mm}

Early multimodal retrievers were largely \emph{TextRAG}-style: they transformed multimodal components into textual surrogates via OCR, captioning, or serialization, enabling mature text retrieval pipelines but inevitably discarding vision-specific cues~\cite{16_mmmulhop, 17_unifiedprese, 18_unifying, 19_helios}.
More recently, \emph{VisRAG}-style pipelines unify modalities by rasterizing documents into page- or region-level screenshots and embedding all content in a single \emph{visual} space~\cite{3_visrag, 4_colpali, 5_m3docvqa}.
However, they suffer from two key limitations: (i) \textit{fixed granularity}, where large screenshots dilute query-relevant signals with irrelevant context, and (ii) \textit{limited multihop reasoning}, treating pages independently without exploiting structural links~\cite{6_densexretrieval, 7_mixofgran}.

Closest to our work, \textsc{LILaC} represents a multimodal document corpus as a layered graph structure and performs structure-aware retrieval through vector-embedding-based graph traversal~\cite{1_lilac}.
It builds a component graph linking coarse nodes (paragraphs, tables, images) and fine-grained subcomponents, using edges to represent both hierarchical containment and navigational relations.
At query time, \textsc{LILaC} performs an edge-wise beam search driven by late-interaction scores between subqueries and nodes.
While effective, this pipeline operates under a rigid, pre-defined plan: it relies on a uniform similarity metric for traversal, overlooking hop-specific semantics, and executes a linear expansion strategy.

\vspace{-2mm}
\subsection{Graph Retrieval Methods}
\vspace{-1mm}

Graph-based retrieval has been extensively studied in knowledge graph QA, where systems traverse graphs curated with \emph{typed} and \emph{semantically meaningful} relations~\cite{13_reasonpath, 24_hierargraph, tog, rog, gog}.
However, multimodal \emph{document} graphs differ fundamentally: edges are primarily \emph{navigational} (e.g., hyperlinks) rather than semantic predicates. 
Consequently, a retriever cannot treat traversal as simple path-finding over valid facts; it must perform online interpretation to resolve what a navigational link implies for the current query context~\cite{1_lilac}.
Thus, KG methods struggle with the adaptive capabilities to navigate large-scale, non-edge-labeled graphs where semantic resolution is needed~\cite{16_mmmulhop, 27_mmapg}.

\vspace{-2mm}
\subsection{Agentic Retrieval Methods}
\vspace{-1mm}

A growing line of work treats retrieval as an iterative decision process interleaved with reasoning, rather than a single-shot nearest-neighbor lookup.
\textsc{IRCoT} shows that multi-step questions benefit from repeatedly generating intermediate sub-questions and retrieving evidence for each step~\cite{9_ircot}.
\textsc{ReAct} formalizes a general reasoning-and-acting loop, motivating RAG controllers that plan retrieval actions based on intermediate observations~\cite{11_react}.

Despite this, most operate on \emph{flat} indices, lacking the structural awareness to navigate inter-document links. 
Furthermore, their error correction is typically limited to query rewriting rather than history-aware backtracking. 
While systems like \textsc{Doc-React} and \textsc{MARA} apply agents to multimodal documents, they target single-document or small-scale contexts~\cite{33_docreact, 35_mara}. 
They do not address the open-domain challenge of traversing vast, interconnected graphs where utilizing failure feedback is critical for routing optimization.

\vspace{-1mm}
\section{Multimodal Document Retrieval}
\vspace{-1mm}
\label{sec:probdef}

We study \emph{open-domain multimodal document retrieval} to find relevant components from a large multimodal corpus on a natural language query.
In this study, we follow the setup of graph-based retrieval approaches~\cite{1_lilac}, which have shown promising performances over existing naive approaches.

\begin{figure}[t]
  \centering
  \includegraphics[width=\linewidth]{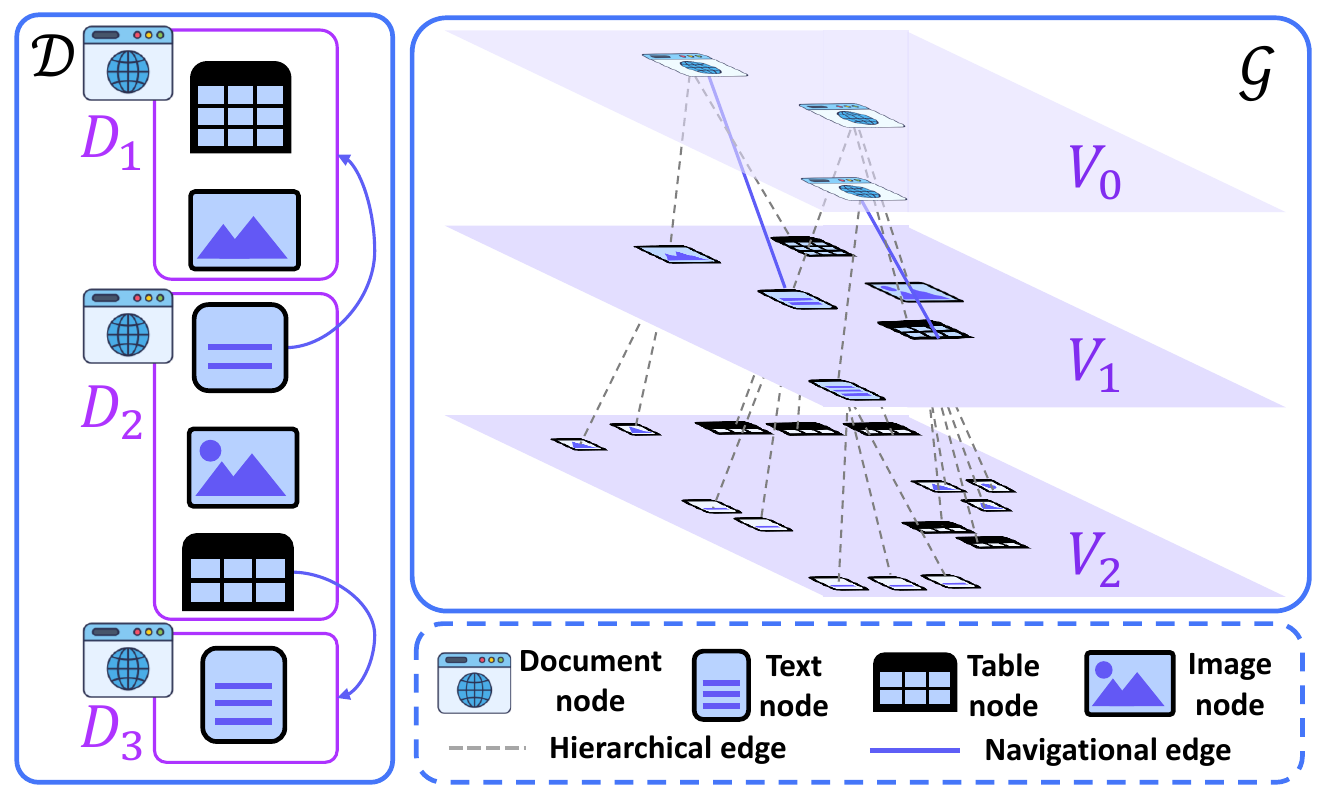}
  \caption{
    Visualization of an example corpus $\mathcal{D}$ and its corresponding layered component graph $\mathcal{G}$.
    }
  \label{fig:preliminaries_example}
  \vspace{-6mm}
\end{figure}

\vspace{-1mm}
\subsection{Problem Setting}
\vspace{-1mm}
\noindent\textbf{Corpus and components.}
For the source of retrieval, a corpus $\mathcal{D}=$$\{D_1,$$\ldots,$$D_{k_{doc}}\}$ is a set of documents.
Each document $D_j$ includes an ordered list of multi-modal components $D_j=[C_{j,1},$ $\ldots,$ $C_{j,k_j}]$, where the global component pool is defined as $\mathcal{C}$ $=$ $\bigcup_{j=1}^{k_{doc}}\{$ $C_{j,1},$ $\ldots,$ $C_{j,k_j}\}$.
The modality of each component $C\in\mathcal{C}$ can be a paragraph $P$, a table $T$, or an image $I$ as shown in Figure~\ref{fig:preliminaries_example}.

\vspace{-1mm}
\noindent\textbf{Navigational links.}
We assume a link signal $\mathcal{L}$ capturing navigational associations (e.g., hyperlinks and cross-document pointers), modeled as $\mathcal{L}:\mathcal{C}\rightarrow\mathcal{D}$.

\vspace{-1mm}
\noindent\textbf{Multimodal retrieval task.}
Given $Q$, $\mathcal{D}$, and $\mathcal{L}$, the retriever ranks components in $\mathcal{C}$ and returns $\mathcal{C}_{R}=[C_{R_1},\ldots,C_{R_n}]$.
Let $\mathcal{C}_{gt}(Q)=\{C_{gt_1},\ldots,C_{gt_r}\}$ be the ground-truth relevant set; the goal is to rank its elements in $\mathcal{R}$, ideally near the top.

\vspace{-1mm}
\subsection{Layered Component Graph}
\label{sec:lcg}
\vspace{-1mm}

We incorporate a layered component graph~\cite{1_lilac} to support an effective \emph{coarse-to-fine} retrieval across documents and their components.
To efficiently retrieve relevant documents and evidence for a given query, open-domain retrieval requires to plan both which documents to visit and which components to read.
Thus, we adopt the three-layered component graph to effectively represent documents, components, subcomponents, and their complex relations.
Figure~\ref{fig:preliminaries_example} shows an example graph $\mathcal{G}$.

\vspace{-1mm}
\noindent\textbf{Nodes and Layers.}
Let $\mathcal{G}=(\mathcal{V},\mathcal{E})$ denote the graph. 
Using the definitions from Section~\ref{sec:probdef}, we construct a three-layered hierarchy $\mathcal{V} = V_0 \cup V_1 \cup V_2$:

\vspace{-2mm}
\squishlist
    \item \textit{Layer 0 (Documents) $V_0$}
    \item \textit{Layer 1 (Components) $V_1$}: paragraphs/tables/images
    \item \textit{Layer 2 (Subcomponents) $V_2$}: sentences, table rows, or visual objects
\squishend
\vspace{-2mm}
For nodes in $V_1\cup V_2$, we save the raw multimodal content.
For nodes in $V_0$, we save a short textual summary of each document to avoid long inputs while preserving high-level semantics for global routing.

Different from LILAC~\cite{1_lilac}, which uses two layers only with components and subcomponents, we add an explicit \emph{Document Layer} to store a concise \emph{textual summary} per document node.
Adding document layer enables early pruning before descending to fine-grained evidence and can significantly improve the retrieval performance.

\vspace{-1mm}
\noindent\textbf{Hierarchical Edges.}
These edges represent the ``contains'' relationship, allowing the agent to drill down from coarse to fine granularity.
\vspace{-2mm}
\squishlist
    \item Edges $(D_j, C_{j,i})$ for all components $C_{j,i} \in D_j$.
    \item Edges linking a component to its extracted subcomponents. 
\squishend
\vspace{-1mm}

\vspace{-1mm}
\noindent\textbf{Navigational Edges.}
These edges capture explicit navigational paths across the corpus, allowing the retriever to transition between different document contexts based on the link signal $\mathcal{L}$.
Using the link signal $\mathcal{L}$, we generate an edge $(C,D_k)$ if $\mathcal{L}(C)=D_k$.

Figure~\ref{fig:preliminaries_example} illustrates the two edge types: dotted lines express hierarchical edges, and blue lines express navigational edges.

\begin{figure*}[t]
  \centering
  \includegraphics[width=1.00\linewidth]{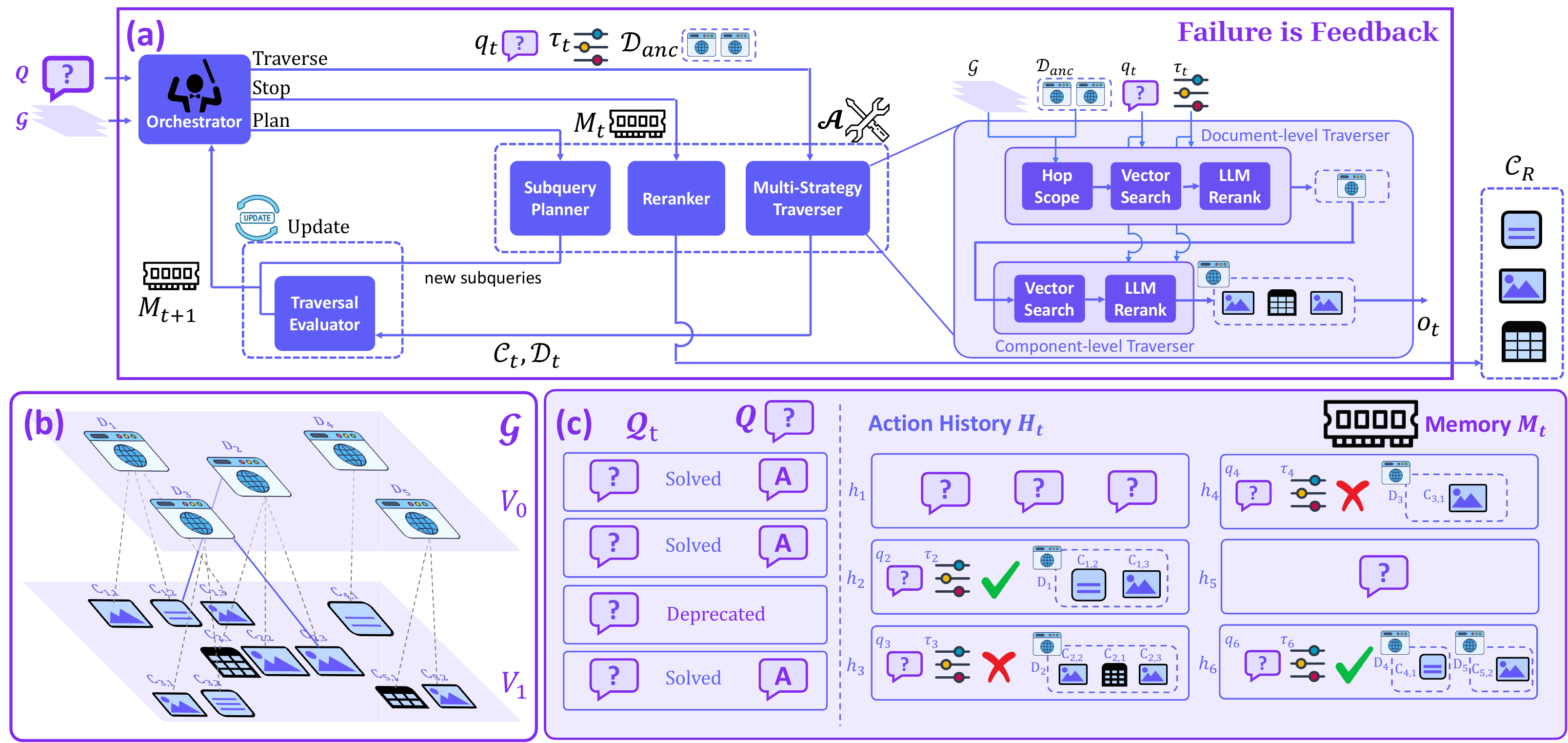}
  \caption{
    Overview of \textsc{\OurFullName}.
    (a) High-level orchestration loop and action interactions, labeled using the notation from the problem formulation (\textsection~\ref{sec:probform}).
    (b) An example layered component graph $\mathcal{G}$. 
    Note that only the document and component nodes are shown for brevity.
    (c) An example memory $M_t$ of a traversal over $\mathcal{G}$.
    }
    \vspace{-4mm}
  \label{fig:overview}
\end{figure*}

\vspace{-1mm}
\section{\textsc{\OurFullName}}
\label{sec:method}
\vspace{-1mm}

{\color{red}

}

We propose \textsc{\OurFullName} (\textsc{\Ours}), an LLM-driven agentic retriever that traverses the layered component graph $\mathcal{G}$. 
Instead of executing a static, pre-defined traversal plan, \textsc{\Ours} formulates retrieval as a \emph{sequential decision process}: 
an Orchestrator iteratively chooses to 
(i) \textsc{Traverse} from selected anchors under a strategy that explicitly trades off accuracy and cost, 
(ii) \textsc{Plan} by revising or expanding subqueries as the information need evolves, or 
(iii) \textsc{Stop} and invoke a final \textsc{Reranker} over all accumulated candidates.
A structured traversal memory records retrieved evidence together with explicit success/failure outcomes, operationalizing our core principle that \emph{failure is feedback}: 
the Orchestrator escalates to stronger (but costlier) reasoning for traversal when lightweight hops are ambiguous or fail, and performs history-aware backtracking by re-anchoring to more promising prior contexts while avoiding previously failed routing patterns.
We start by formalizing the sequential decision process, then explain the details for each agent that comprise the process.

\vspace{-1mm}
\subsection{Retrieval as a Sequential Decision Process}
\label{sec:probform}
\vspace{-1mm}

We formulate open-domain multimodal retrieval as a \textit{sequential decision process} over the layered component graph $\mathcal{G}$.
Given a query $Q$, the agent iteratively traverses $\mathcal{G}$ to output an ordered list of relevant components.
We formalize the process as $\langle \mathcal{S}, \mathcal{A}, \mathcal{T}\rangle$, where the state is the agent's information state $s_t$; 
executing $a_t$ yields observation $o_t$, which is integrated into $s_{t+1}$ by the transition.

\noindent\textbf{State ($\mathcal{S}$).}
We represent the information state as a structured \textit{memory} $s_t = M_t$.
$M_t$ is basically a trajectory log capturing decisions and outcomes: $M_t = (Q, \mathcal{Q}_t, H_t)$.
\vspace{-2mm}
\squishlist
    \item \textit{Original Query ($Q$):} The user's initial input.

    \item \textit{Subquery List ($\mathcal{Q}_t$):} a list of decomposed query serving as a retrieval plan.

    \item \textit{Action History ($H_t$):}
    an ordered sequence $[h_1,$ $\ldots,$ $h_t]$.
    Each record $h_k$ serves as a log of each action.
\squishend
\vspace{-2mm}


\noindent\textbf{Action ($\mathcal{A}$).}
An action $a_t \in \mathcal{A}$ is a structured tool call selected by the orchestrator given $s_t$:
\vspace{-2mm}
\squishlist

    \item $\textsc{Traverse}(\mathcal{D}_{anc}, q_t, \tau_t)$ executes one retrieval hop for subquery $q_t$ with strategy $\tau_t$ from an anchor document set $\mathcal{D}_{anc}$.
    
    \item $\textsc{Plan}(M_t)$ generates updated subqueries needed to solve $Q$, consulting both $\mathcal{Q}_t$ and $H_t$.

    \item $\textsc{Stop}$ terminates the overall process.
    It applies a final \textsc{Rerank} module to all components stored in memory $M_t$ and returns a ranked list $\mathcal{C}_R$ based on their relevance to the original query $Q$.

\squishend
\vspace{-2mm}

\noindent\textbf{Observation ($o_t$).}
$o_t$ contains the \emph{new} outputs produced by the action at step $t$.
For \textsc{Traverse}, the observation is the traversed documents/components and if the traversal was successful or not.
For \textsc{Plan}, the observation is the updated list of subqueries.

\noindent\textbf{Transition ($\mathcal{T}$).}
The transition process updates the state $s_t$ using the observation $o_t$ to generate the next state $s_{t+1}$.
Specifically, it appends the observation to the action history $H_t$.
It appends the newly generated subqueries to $\mathcal{Q}_t$.

    
    
    

\vspace{-1mm}
\subsection{Action Design}
\vspace{-1mm}


In this subsection, we detail the LLM-powered agents that implement each action in our sequential decision process defined in \textsection~\ref{sec:probform}.
Crucially, \textsc{\Ours}'s actions are designed to make two principles \emph{operational}:
(i) \emph{economically-rational control} via a portfolio of traversal strategies that can be escalated on demand, and
(ii) \emph{failure-as-feedback} via explicit success/failure signals and history-aware re-anchoring for robust multihop navigation in noisy document graphs.

\noindent\textbf{(1) Multi-strategy Traverser.}
It executes the $\textsc{Traverse}(\mathcal{D}_{anc}, q_t, \tau_t)$ action on the layered graph $\mathcal{G}$, aiming to find subquery $q_t$-relevant components and documents, anchoring from the documents $\mathcal{D}_{anc}$.
It consists of two stages: 
(i) a \emph{document-level traverser} that selects a small set of candidate documents using document-node summaries (Layer $V_0$), and
(ii) a \emph{component-level traverser} that identifies $q_t$-relevant components among the candidates by searching their children nodes.

Crucially, the Traverser is a \emph{configurable strategy engine}: the strategy tuple $\tau_t$ specifies retrieval behaviors that trade off accuracy and efficiency.
\textbf{Strategy tuple.}
$\tau_t$ controls traversal along three dimensions:
\squishlist

    \item \textit{Hop Scope (Global vs.\ Local).}
    \emph{Local Hop} restricts candidate documents to the direct neighbors of the anchor document set $\mathcal{D}_{anc}$ via the children's navigation edges;
    \emph{Global Hop} considers the full document corpus to escape local neighborhoods when evidence is dispersed.
    It performs a dense retrieval over layer $V_0$'s document summaries to search the most query-relevant document set.
    
    \item \textit{Vector Scoring Granularity.}
    It configures the layer at which vector similarity is computed.
    Intuitively, coarse-grained subqueries are often best matched at the document/component level, while fine-grained subqueries benefit from scoring at the subcomponent level.
    The document-level traverser uses the granularities of $g\in\{0,1,2\}$ and the component-level uses $g\in\{1,2\}$.
    Let $\mathrm{Desc}_g(v)$ be descendants of $v$ at Layer $V_g$ (including $v$ if already in $V_g$). 
    We score a node $v$ by
    \begin{equation*}
        \textsc{Score}_{\mathrm{vec}}(q_t, v; g)
        \;=\;
        \max_{u \in \mathrm{Desc}_g(v)} sim(q_t, x(u))
    \end{equation*}
    
    \item \textit{LLM-Reasoning.}
    This option specifies whether to use LLM-based reasoning to accurately rerank components with $q_t$ beyond vector scores, supporting modes of $\{\texttt{None},\ \texttt{Component-only},\ \texttt{Both}\}$.
    To control LLM inference cost, we pre-filter top-$k$ candidates by $\textsc{Score}_{\mathrm{vec}}$ and then pass their contents to the LLM for reranking.
    
\squishend
The traversal outputs a document set $\mathcal{D}_t$ and a component set $\mathcal{C}_t$, which are recorded in the action history $H_t$.

\vspace{-1mm}
\noindent\textbf{(2) Subquery Planner.}
The Planner implements $\textsc{Plan}(M_t)$ and generates new subqueries to address the remaining information needs for answering $Q$, conditioning on the current subquery statuses $\mathcal{Q}_t$ and the retrieved evidence stored in $H_t$.
Invoking \textsc{Plan} allows the agent to revise its plan using the newly acquired evidence in the cases where initial subquery list is suboptimal.
Newly generated subqueries are appended to the subquery list $\mathcal{Q}$ in memory $M_t$; importantly, we keep earlier subqueries rather than overwriting them, so the system retains a trace of what was tried and where the plan went off track.

\vspace{-1mm}
\noindent\textbf{(3) Traversal Evaluator.}
The Evaluator produces part of the observation $o_t$ of \textsc{Traverse} by assessing whether the traversal outcome $\mathcal{C}_t$ is useful for the target subquery $q_t$.
It provides two outputs.
First, it judges whether the retrieved components can answer $q_t$; this success/failure signal is stored in the action history $H_t$ together with the traversal results.
Second, it checks whether the retrieved content can resolve any remaining items in the subquery list $\mathcal{Q}_t$, and if so, extracts tentative answers and updates $\mathcal{Q}_t$ accordingly.
These logged outcomes later guide backtracking decisions, helping the Orchestrator distinguish promising contexts from unproductive ones.

\vspace{-1mm}
\noindent\textbf{(4) Reranker.}
The Reranker is invoked by $\textsc{Stop}$ to produce the final ranked list of components $\mathcal{R}$ from the accumulated memory $M_t$.
Concretely, it aggregates all candidate components stored in memory and assigns each a final relevance score with respect to the \emph{original} query $Q$, then returns $\mathcal{C}_R$ by selecting the top-$k$ components.

\vspace{-1mm}
\subsection{Orchestrator}
\label{sec:tools}
\vspace{-1mm}

The Orchestrator is the central LLM-driven controller that implements the policy over actions, selecting $a_t$ given the current information state $M_t$.
At each iteration, it chooses among $\textsc{Traverse}$, $\textsc{Plan}$, and $\textsc{Stop}$ to maximize retrieval accuracy while remaining efficient, treating \emph{failure as feedback}.

Beyond selecting actions, the \textsc{\Ours} Orchestrator actively \emph{manages plan errors}.
Rather than using $q_t$ as a direct copy from the subquery list, it treats the list as a scaffold and synthesizes a task-specific target by rewriting, refining, or composing subqueries based on the current evidence and unresolved constraints.
In particular, it can use newly retrieved components to resolve missing information from earlier subqueries, or fuse multiple planned subqueries into a single sharper retrieval objective when they are interdependent.
When the subquery list becomes clearly unreliable (e.g., overly underspecified, drifting, or repeatedly unproductive), the Orchestrator can invoke $\textsc{Plan}$ at any time to regenerate a better set of subqueries conditioned on $M_t$, while preserving prior subqueries as a trace of what was attempted.

The Orchestrator also enables \emph{history-aware backtracking via re-anchoring}.
Instead of always hopping from the most recent set of documents, it chooses an anchor document set $\mathcal{D}_{anc}$ that best matches the newly formed $q_t$, potentially restarting from an earlier successful context.
To do so, it consults the action history $H_t$, leveraging (i) summaries of previously retrieved documents and (ii) the Traversal Evaluator's success/failure outcomes to identify the most promising region to resume from and to avoid repeating failed routing patterns.
This re-anchoring makes multihop navigation more resilient to dead ends and spurious cues.
In addition, the Orchestrator explicitly reasons about efficiency through the strategy tuple: 
for each new $q_t$, it predicts an economical traversal configuration $\tau_t$ to try first.
If such a hop fails, it can retry from the same anchor (or a better re-anchored set) with an escalated, more accurate $\tau_t$ (e.g., finer granularity, leveraging LLM reranking, or using global scope).
This closed-loop control reduces failures caused by insufficient reasoning while keeping computation focused where it is most needed.

\begin{table*}[htbp]
    \caption{Retrieval accuracy (Recall and MRR) of \textsc{\Ours} and its competitors on three benchmarks. R@k, M@k indicates recall at $k$ and MRR at $k$, respectively. The best score in each column is in \textbf{bold}. }
    \vspace{-3.5mm}
    \label{tab:retrieval_performance}
    \begin{center}
    \begin{small}
    \scalebox{1.00}{
    \renewcommand{\arraystretch}{0.95}
    \resizebox{1.0\textwidth}{!}{
    \begin{tabular}{l|rrrrr|rrrrr|rrrrr}
        \toprule
        \multirow{2}{*}{Algorithm} & 
        \multicolumn{5}{c|}{\MultimodalQA} & 
        \multicolumn{5}{c|}{\MMCoQA} & 
        \multicolumn{5}{c}{\WebQA} \\
        \cmidrule(lr){2-6} \cmidrule(lr){7-11} \cmidrule(lr){12-16}
        & R@1 & R@2 & R@5 & R@10 & M@10 
        & R@1 & R@2 & R@5 & R@10 & M@10 
        & R@1 & R@2 & R@5 & R@10 & M@10 \\
        \midrule
        
        \textsc{NV-Embed-v2} & 
            26.42 & 37.63 & 52.08 & 61.45 & 68.13 & 
            18.22 & 28.46 & 40.22 & 48.19 & 41.97 & 
            22.18 & 31.38 & 39.80 & 51.19 & 55.81 \\
            
        \textsc{VisRAG}  & 
            34.19 & 42.12 & 53.38 & 56.91 & 57.88 &
            19.57 & 27.53 & 33.59 & 37.31 & 30.01 &
            25.58 & 41.46 & 46.10 & 48.75 & 50.60 \\
            
        \textsc{ColPali}     & 
            38.38 & 52.61 & 61.73 & 63.95 & 67.65 &
            31.11 & 41.97 & 46.89 & 51.27 & 43.13 &
            35.21 & 46.90 & 53.37 & 57.78 & 56.90 \\
            
        \textsc{LILaC}       & 
            33.59 & 50.59 & 65.13 & 72.23 & 79.12 & 
            25.25 & 38.13 & 51.02 & 60.86 & 53.36 & 
            32.67 & 45.35 & 57.77 & 64.23 & 77.87 \\
            
        \textsc{IRCoT}       & 
            39.56 & 55.09 & 69.87 & 74.97 & 82.24 & 
            43.03 & 53.69 & 62.70 & 64.62 & 62.84 &
            40.85 & 56.73 & 68.78 & 73.72 & 83.15 \\
            
        \textsc{\Ours}         & 
            \textbf{42.17} & \textbf{57.58} & \textbf{74.88} & \textbf{85.82} & \textbf{86.82} & 
            \textbf{46.31} & \textbf{58.40} & \textbf{69.47} & \textbf{75.17} & \textbf{74.88} & 
            \textbf{43.91} & \textbf{61.78} & \textbf{74.83} & \textbf{80.79} & \textbf{87.77} \\
        
        \bottomrule
    \end{tabular}
    }
    \renewcommand{\arraystretch}{1.0}
    }
    \end{small}
    \end{center}
    \vspace{-6mm}
\end{table*}

\begin{table}[!t]
    \caption{End-to-end QA accuracy (EM and F1) of \textsc{\Ours} and its competitors for the three bnechmarks. The best score in each column is in \textbf{bold}.}
    \vspace{-3.5mm}
    \label{tab:qa_em_f1}
    \begin{center}
    \begin{small}
    \scalebox{0.9}{
    \renewcommand{\arraystretch}{0.95}
    \resizebox{1.1\linewidth}{!}{
    \begin{tabular}{lcccccc}
        \toprule
        
        \multirow{2}{*}{Algorithm} &
        \multicolumn{2}{c}{\tiny \MultimodalQA} &
        \multicolumn{2}{c}{\small \MMCoQA} &
        \multicolumn{2}{c}{\small \WebQA} \\
        \cmidrule(lr){2-3}\cmidrule(lr){4-5}\cmidrule(lr){6-7}
        & EM & F1 & EM & F1 & EM & F1 \\
        \midrule
        
        \textsc{NVEmbed-v2}                 & 53.80 & 61.15 & 39.20 & 47.73 & 51.82 & 59.88 \\
        \textsc{VisRAG}                 & 51.40 & 61.72 & 35.20 & 42.66 & 46.53 & 55.19 \\
        \textsc{ColPali}                    & 52.82 & 63.14 & 40.73 & 47.46 & 49.87 & 57.78 \\
        \textsc{LILaC}                      & 57.78 & 63.98 & 42.57 & 51.14 & 53.38 & 60.71 \\
        \textsc{IRCoT}                      & 60.32 & 68.09 & 47.21 & 57.71 & 57.03 & 65.88 \\
        \textsc{\Ours}                      & \textbf{65.15} & \textbf{70.47} & \textbf{51.11} & \textbf{ 62.42} & \textbf{62.63} & \textbf{72.47} \\
        \bottomrule
        
    \end{tabular}
    }
    \renewcommand{\arraystretch}{1.0}
    }
    \end{small}
    \end{center}
    \vspace{-6mm}
\end{table}

\vspace{-2mm}
\section{Experiments}
\label{sec:experiments}
\vspace{-1mm}

\begin{table*}[!t]
    \caption{Efficiency (Time, \# LLM Calls and API Usage) comparison of \textsc{\Ours} and its competitors for the three benchmarks.}
    \vspace{-3.5mm}
    \label{tab:retrieval_efficiency}
    \begin{center}
    \begin{small}
    \scalebox{0.82}{
    \renewcommand{\arraystretch}{0.95}
    \resizebox{1.2\textwidth}{!}{
    \begin{tabular}{llrrrrrrrr}
    
        \toprule
        \multirow{2}{*}{Dataset} & \multirow{2}{*}{Algorithm} & 
        \multicolumn{4}{c}{Time (ms)} & \multirow{2}{*}{\# LLM Calls} &
        \multicolumn{3}{c}{API Usage} \\
        \cmidrule(lr){3-6}\cmidrule(lr){8-10}
        & & Total & LLM & Vector Search & Embedding & & \# Input Toks & \# Output Toks & \$ \\
        \midrule
        
        \multirow{5}{*}{\MultimodalQA}
        & \textsc{VisRAG}  & 371    & 0      & 218   & 153 & 0.00 & 0      & 0     & 0.0000 \\
        & \textsc{ColPali} & 9,849  & 0      & 9,210 & 639 & 0.00 & 0      & 0     & 0.0000 \\
        & \textsc{LILaC}   & 19,528 & 15,943 & 3,153 & 432 & 1.00 & 1,165  & 1,119 & 0.0115 \\
        & \textsc{IRCoT}   & 95,744 & 95,140 & 119   & 484 & 5.44 & 41,943 & 3,296 & 0.0752 \\
        & \textsc{\Ours}   & 114,591& 113,642& 538   & 411 & 7.92 & 48,863 & 4,785 & 0.1109 \\
        \midrule
        
        \multirow{5}{*}{\MMCoQA}
        & \textsc{VisRAG}  & 362    & 0      & 215   & 147 & 0.00 & 0      & 0     & 0.0000 \\
        & \textsc{ColPali} & 1,836  & 0      & 1,173 & 663 & 0.00 & 0      & 0     & 0.0000 \\
        & \textsc{LiLaC}   & 20,244 & 16,879 & 2,977 & 388 & 1.00 & 1,160  & 1,041 & 0.0107 \\
        & \textsc{IRCoT}   & 97,816 & 97,220 & 116   & 479 & 5.61 & 52,286 & 3,360 & 0.0753 \\
        & \textsc{\Ours}   & 105,355& 104,630& 366   & 359 & 7.31 & 53,149 & 4,151 & 0.1037 \\ 
        \midrule
        
        \multirow{5}{*}{\WebQA}
        & \textsc{VisRAG}  & 386    & 0      & 225   & 161 & 0.00 & 0      & 0     & 0.0000 \\
        & \textsc{ColPali} & 7,919  & 0      & 7,298 & 621 & 0.00 & 0      & 0     & 0.0000 \\
        & \textsc{LILaC}   & 19,187 & 14,822 & 3,782 & 583 & 1.00 & 1,162  & 737   & 0.0077 \\
        & \textsc{IRCoT}   & 173,815& 173,011& 152   & 651 & 6.94 & 65,167 & 3,913 & 0.1082 \\
        & \textsc{\Ours}   & 128,748& 127,664& 581   & 505 & 8.61 & 64,683 & 5,159 & 0.1278 \\
        \bottomrule
        
    \end{tabular}
    }
    \renewcommand{\arraystretch}{1.0}
    }
    \end{small}
    \end{center}
    \vspace{-4mm}
\end{table*}

\begin{table*}[htbp]
    \caption{Ablation study analyzing retrieval accuracy and efficiency of different \textsc{\Ours} variants.}
    \vspace{-3.5mm}
    \label{tab:ablation_study}
    \begin{center}
    \begin{small}
    \scalebox{0.82}{
    \renewcommand{\arraystretch}{0.95}
    \resizebox{1.2\textwidth}{!}{
    \begin{tabular}{llccrcccc}
    
        \toprule
        \multirow{2}{*}{Dataset} & \multirow{2}{*}{Variation} &
        \multicolumn{2}{c}{Retrieval Accuracy} &
        \multicolumn{5}{c}{Efficiency} \\
        \cmidrule(lr){3-4}\cmidrule(lr){5-9}
        & & R@10 & MRR@10 & Time (ms) & \# LLM Calls & \# Input Toks & \# Output Toks & \$ \\
        \midrule
        
        \multirow{6}{*}{\footnotesize \MultimodalQA}
        & \textsc{\OurFullName}                        & 85.82 & 86.82 & 
                                                    117,651 & 7.97 & 49,153 & 4,812 & 0.1111 \\
                                                    
        & w/o Backtracking Orchestration        & 79.26 & 84.31 &
                                                    211,951 & 12.53 & 78,864 & 6,113 & 0.1502 \\
                                                    
        & w/o LLM Reasoning in traversal agent  & 78.04 & 83.97 & 
                                                    168,780 & 7.64 & 45,079 & 4,203 & 0.0887 \\
                                                    
        & w/o Global Hop                        & 75.79 & 82.56 & 
                                                    85,417 & 7.02 & 51,839 & 4,137 & 0.1039 \\
                                                    
        & w/o Vector Granularity                & 83.16 & 85.43 &
                                                    116,918 & 8.28 & 55,281 & 5,179 & 0.1158 \\
                                                    
        & w/o Subquery Planner                  & 76.2 & 83.77 & 
                                                    66,974 & 5.31 & 35,984 & 3,335 & 0.0765 \\
        \midrule
        
        \multirow{6}{*}{\WebQA}
        & \textsc{\OurFullName}                        & 81.91 & 89.24 & 
                                                    126,567 & 8.55 & 63,971 & 5,103 & 0.1270 \\
                                                    
        & w/o Backtracking Orchestration        & 77.45 & 87.62 & 
                                                    228,934 & 13.77 & 89,244 & 7,214 & 0.1715 \\
                                                    
        & w/o LLM Reasoning in traversal agent  & 78.92 & 88.82 & 
                                                    135,274 & 7.57 & 50,501 & 4,275 & 0.0960 \\
                                                    
        & w/o Global Hop                        & 73.30 & 86.83 & 
                                                    93,115 & 9.05 & 68,483 & 4,082 & 0.1247 \\
                                                    
        & w/o Vector Granularity                & 79.21 & 87.77 &
                                                    132,972 & 8.72 & 66,201 & 5,319 & 0.1290 \\
                                                    
        & w/o Subquery Planner                  & 76.38 & 89.95 & 
                                                    85,583 & 5.40 & 40,721 & 3,358 & 0.0820 \\
        \bottomrule
        
    \end{tabular}
    }
    \renewcommand{\arraystretch}{1.0}
    }
    \end{small}
    \end{center}
    \vspace{-6mm}
\end{table*}

\vspace{-1mm}
\subsection{Experimental Setups}
\vspace{-1mm}

\noindent\textbf{Datasets and evaluation metrics.}
We evaluate open-domain multimodal \emph{component} retrieval and downstream QA on three benchmarks: \textsc{MultimodalQA}~\cite{multimodalqa}, \textsc{MMCoQA}~\cite{mmcoqa}, and \textsc{WebQA}~\cite{webqa}.
Following \textsc{LILaC}~\cite{1_lilac}, we use the URL-annotated setting to reconstruct realistic webpage-style corpora, parsing each page into multimodal components (paragraphs, tables, images).
This yields \MultimodalQA (3,235 pages, avg. $\sim$37 components), \MMCoQA (453 pages, avg. $\sim$32 components), and \WebQA (7,662 pages, avg. $\sim$13 components).
Consistent with prior work~\cite{1_lilac}, we report retrieval Recall@$k$ (R@$k$, $k\in\{1,2,5,10\}$) and MRR@10: R@$k$ checks whether at least one ground-truth component appears in the top-$k$ list, and MRR@10 captures the rank of the first relevant component.
For end-to-end QA, we feed the top-$10$ retrieved components into the same multimodal LLM and report Exact Match (EM) and token-level F1.

\vspace{-1mm}
\noindent\textbf{Compared methods.}
We compare \textsc{\Ours} with strong baselines spanning graph traversal, agentic retrieval, and single-shot indexing.
We include \textsc{LILaC}~\cite{1_lilac}, a layered-graph retriever designed for multi-hop scenarios, and \textsc{IRCoT}~\cite{9_ircot}, an agentic retriever that interleaves retrieval with chain-of-thought reasoning.
Since \textsc{IRCoT} was originally proposed for text-only corpora, we adapt it to our multimodal setting by (i) replacing its retriever with the same multimodal embedder used throughout our experiments and (ii) using the same multimodal LLM for reasoning and generation over multimodal components.
For \emph{VisRAG} approaches, we employ \textsc{VisRAG-Ret}~\cite{3_visrag}, which directly encodes document images via VLMs, and \textsc{ColPali}~\cite{4_colpali}, which uses late-interaction multi-vector embeddings from document images.
We also compare with \textsc{NV-Embed-v2}~\cite{nvembedv2}, a \emph{TextRAG} baseline that embeds textualized components.

\vspace{-1mm}
\noindent\textbf{Model configurations.}
To ensure fair comparison, we standardize backbone models across all methods whenever applicable.
We use \textsc{MM-Embed}~\cite{mmembed} as the unified multimodal embedder, and the \textsc{OpenAI API} (\textsc{gpt-5})~\cite{gpt5} with \texttt{reasoning\_effort = low} as the multimodal LLM for all planning, reasoning, reranking, and generation steps.

\vspace{-1mm}
\subsection{Retrieval Accuracy Comparison}
\label{sec:retrieval_accuracy}
\vspace{-1mm}

We evaluate open-domain multimodal \emph{component} retrieval on \MultimodalQA, \MMCoQA, and \WebQA\ using Recall@$k$ ($k\in\{1,2,5,10\}$) and MRR@10, with results reported in Table~\ref{tab:retrieval_performance}.
\textsc{\Ours} achieves the best performance across all three benchmarks and all reported cutoffs, indicating its overall effectiveness.
Averaged across datasets, \textsc{\Ours} reaches average Recall@10 of 80.26 and average MRR@10 of 83.16, improving over \textsc{LILaC} +22.03\% and +18.60\%, respectively.
Compared with the strongest agentic baseline \textsc{IRCoT}, \textsc{\Ours} gains 12.88\% Recall@10 and 9.31\% MRR@10.
The gap is even larger against single-shot embedding-based retrievers.

We analyze two interesting points.
One is that the largest dataset-specific margin over \textsc{LILaC} appears on \WebQA at higher cutoffs, suggesting that \WebQA more frequently requires \emph{escaping} local neighborhoods to reach dispersed evidence.
This aligns with \WebQA's construction where ground-truth components are not necessarily adjacent or tightly coupled, making global navigation critical.
In contrast, the performance gap over \textsc{IRCoT} is most pronounced on the more explicitly multi-hop benchmarks \MultimodalQA and \MMCoQA, where effectively leveraging the underlying link/structure signal (rather than pure global searching) is essential.
Second, improvements are particularly strong at low-$k$, indicating that \textsc{\Ours} not only increases \emph{coverage} of relevant components but also ranks them substantially earlier.

\vspace{-1mm}
\subsection{End-to-end QA Accuracy Comparison}
\vspace{-1mm}

We measure end-to-end QA performance by feeding the top-$10$ retrieved components into the same multimodal LLM generator for every method, and report EM and token-level F1 in Table~\ref{tab:qa_em_f1}.
\textsc{\Ours} is consistently the best-performing method on all three datasets, achieving average EM/F1 of 59.63/68.45.
Relative to \textsc{LILaC}, \textsc{\Ours} improves EM by +16.37\% and F1 by +16.79\% on average, confirming that more reliable evidence discovery yields better grounded generation.
Compared to \texttt{IRCoT}, \textsc{\Ours} still provides a clear advantage of +8.71\% EM and +7.14\% F1, despite both methods being agentic.
Dataset-wise, the gains are especially visible on \MMCoQA and \WebQA, consistent with Table~\ref{tab:retrieval_performance} where \textsc{\Ours} yields substantially higher top-$k$ retrieval accuracy.

\vspace{-1mm}
\subsection{Algorithm Efficiency}
\vspace{-1mm}

Table~\ref{tab:retrieval_efficiency} reports wall-clock retrieval time (with breakdown into LLM, vector search, and embedding), the number of LLM calls, token usage, and estimated API cost.
As expected, single-shot retrievers (\textsc{VisRAG}, \textsc{ColPali}, \textsc{NV-Embed-v2}) are the fastest and incur no LLM API cost during retrieval.
Among agentic methods, \textsc{\Ours} shows a slightly lower average runtime than \textsc{IRCoT} (116,231 vs.\ 122,458 ms), which is the strongest agentic baseline: 
while \textsc{\Ours} executes more reasoning steps (7.31--8.61 vs.\ 5.44--6.94 LLM calls), it uses fewer input tokens and achieves comparable or lower latency overall.
Notably, \textsc{\Ours} is substantially faster on \WebQA (128,748 vs.\ 173,815 ms; $-25.93\%$), suggesting that structure-aware navigation together with failure-aware re-anchoring reduces unproductive reasoning on large and sparsely connected corpora; 
The efficiency comes with a moderate increase in API usage compared to \textsc{IRCoT} (\$0.10--\$0.13 vs.\ \$0.075--\$0.108 per query), consistent with our higher retrieval/QA accuracy.
Relative to \textsc{LILaC}, \textsc{\Ours} is 5.20--6.71$\times$ slower in wall-clock time and incurs 9.64--16.60$\times$ higher API cost, quantifying the additional budget required by adaptive agentic control.

\vspace{-1mm}
\subsection{Ablation Study}
\vspace{-1mm}

Table~\ref{tab:ablation_study} isolates the contribution of each major design component in \textsc{\Ours} on \MultimodalQA and \WebQA, reporting retrieval accuracy (R@10, MRR@10) alongside efficiency.
We omit \MMCoQA\ from this analysis: because it is conversational, accumulated dialogue history can introduce confounding factors (e.g., varying context length and carryover information) that may blur the impact of individual retrieval modules.
We run all ablations on a representative 10\% subset of each dataset due to OpenAI API costs.
(i) \textit{History-aware backtracking orchestration.} 
Removing backtracking consistently hurts both \emph{effectiveness} and \emph{efficiency}: R@10 drops by 4.46--6.56, while runtime increases by $\sim$80\% and LLM calls rise by 57--61\%.
This highlights that backtracking prevents wasted hops by adapting effort only when needed.
It also improves accuracy by returning to more promising anchors and narrowing candidates to the right neighborhood, rather than repeatedly exploring uninformative branches.
(ii) \textit{LLM reasoning inside the traversal agent.} 
Disabling LLM reasoning during traversal substantially degrades retrieval and can even \emph{slow down} the search: on \MultimodalQA, R@10 drops by 7.78 and runtime increases by 43\%, despite lower per-step cost.
We reason that additional replanning and extra iterations were triggered, as traversal is more likely to take ambiguous or unproductive hops without LLM reasoning.
(iii) \textit{Global hop.}
Global hops are crucial for escaping local neighborhoods.
When disabled, R@10 suffers the largest drop (10.03 on \MultimodalQA; 8.61 on \WebQA), even though the variant becomes faster.
This indicates that neighbor-based traversal alone is insufficient: some questions require jumping across distant regions of the corpus, including multi-hop paths and cases where relevant evidence is not directly linked.
(iv) \textit{Adaptive vector-search granularity.}
Removing granularity adaptation yields consistent but smaller drops (about 2.66--2.70 R@10) with minimal efficiency change.
(v) \textit{Subquery planner.}
emoving replanning reduces LLM calls and cost substantially (e.g., $-33\%$ to $-37\%$ calls), but causes large recall drops (9.62 on \MultimodalQA; 5.53 on \WebQA).
Without the option to revise the plan, the Orchestrator is more likely to terminate early once progress stalls, which simultaneously lowers cost and recall.
This underscores the importance of correcting early planning mistakes through evidence-conditioned replanning.

\vspace{-2mm}
\section{Conclusion}
\vspace{-1.5mm}

We propose \textsc{\OurFullName} (\textsc{\Ours}), an agentic multimodal retriever that models graph traversal as a \textit{sequential decision process} over an explicit information state.
\textsc{\Ours} maintains a structured memory of subqueries and action history, transforming failures into actionable signals rather than dead ends. 
Building on this, the Orchestrator makes retrieval \emph{economically rational}: it dynamically chooses when to traverse, replan, or stop, and performs cost-aware strategy escalation—starting from low-cost matching and selectively pusuing accuracy over efficiency when ambiguity or failure justifies the added computation.
Finally, \textsc{\Ours} introduces history-aware backtracking via re-anchoring, using history to resume from more promising prior contexts. 
Extensive experiments on \textsc{MultimodalQA}, \textsc{MMCoQA}, and \textsc{WebQA} demonstrate state-of-the-art accuracy on retrieval and downstream QA, validating the effectiveness of failure-aware traversal.%

\newpage
\bibliography{main}
\bibliographystyle{icml2026}

\appendix
\onecolumn

\section{Limitations}

\textsc{\Ours} incurs higher wall-clock latency and API cost than lightweight or fixed-call traversal methods such as \textsc{LILaC}.
Although our cost-aware strategy escalation mitigates unnecessary computation, multi-step orchestration and occasional LLM-based reranking remain as bottlenecks. 
Our method assumes that each webpage/document is already parsed into a clean set of multimodal components and that navigational signals are reliably extracted.
While we focus on URL-annotated corpora with structured navigation and component graphs, truly open-web deployment may involve noisier pages, weaker link signals, dynamic content, and heterogeneous layouts.
An important direction for future work is to develop retrieval pipelines that are robust to diverse webpage structures and can jointly learn or adapt the parsing, graph construction, and retrieval policy so that the approach generalizes more seamlessly to arbitrary web content.

\section{Model Details}
\label{sec:appendix_model_details}

\textbf{(Multimodal) Large language models:}
\squishlist
    \item \texttt{Open-AI ChatGPT5}
\squishend

\noindent \textbf{Text embedders}
\squishlist
    \item \texttt{NV-Embed-v2}: 7.85B parameters
\squishend

\noindent \textbf{Cross-modal embedders:}
\squishlist
    \item \texttt{ColPali}: 3B parameters
    \item \texttt{VisRAG}: 3.43B parameters
\squishend

\noindent \textbf{Multimodal embedders:}
\squishlist
    \item \texttt{MM-Embed}: 8.18B parameters
\squishend

\section{Experimental Details}
\label{appendix:exp_details}

\subsection{Hardware and Software Settings}
\label{appendix:hardware_software_settings}
All experiments were conducted on a Linux server equipped with an Intel Xeon Gold 6230 CPU @ 2.10~GHz, 1~TB of RAM, and four NVIDIA RTX~A6000 GPUs, running Ubuntu~22.04.3~LTS.

\subsection{Implementation Details}
\label{sec:implementation_details}

Our main hyperparameter is the vector-search shortlist size $k$ used by the Multi-strategy Traverser at both the document-level and component-level stages.
Unless otherwise noted, we set $k{=}30$ for all experiments.

\subsection{Benchmark Details}

\MultimodalQA: We use the extended version of \texttt{MultimodalQA}, following the augmentation procedure introduced in M3DocRAG~\cite{m3docrag}. 
It spans diverse document modalities (text, images, and tables) and is designed to stress multi-hop reasoning over multiple documents. 
The evaluation split contains 2,441 questions grounded in 3235 webpages.

\MMCoQA: We use an extended variant of \texttt{MMCoQA} that moves beyond the original distractor-only setting to evaluate conversational, multi-turn multimodal QA. 
The benchmark consists of coherent dialogue sessions in which later questions depend on earlier context and require aggregating evidence across text, images, and tables. 
It includes 5,753 questions grouped into 1,179 conversations, with a corpus of 218,285 text passages, 10,042 tables, and 57,058 images.

\WebQA:
\WebQA\ is a Wikipedia-based multimodal QA benchmark with 4,966 questions over 7,662 documents.
Because the original answers are often verbose, we rewrite them into concise references using the ChatGPT-5 OpenAI API with the prompt below.

\begin{promptbox}{Answer Concisification Prompt for \WebQA}
You are an assistant that extracts concise answers from an Original Answer.

Task:
Given a Question, its Question Category (Qcate), and an Original Answer, extract a concise version of the answer.

Category hints:
- YesNo: respond only "Yes" or "No" matching the polarity of the Original Answer.
- text: return the minimal noun phrase/name that answers the question.
- choose: return only the chosen option or label.
- number: return the numeric value (and unit if present) without extra words.
- color: return the color term(s) only.
- shape: return the shape descriptor only.
- Others: follow the general concise rules below.

Rules:
- Output ONLY the concise answer text (no extra words, no labels, no punctuation-only output).
- Keep the minimum span that directly answers the Question.
- Prefer a single word when possible.
- If the question asks what object/thing, output the object noun phrase only (e.g., fountain).
- If the question asks for a choice/comparison attribute (e.g., taller or shorter, happy or upset, or similar), output only the chosen option word from the answer (e.g., "taller", "upset").
- If the Original Answer is verbose by repeating or paraphrasing words/phrases already present in the Question, do NOT copy those repeated Question words into the concise answer; extract only the new, directly-question-answering information (If those repeated words are necessary for answering the question, then you may include them).
- Preserve the original casing/pluralization as used in the Original Answer (e.g., "Circles").
- Do not include locations, explanations, or surrounding context unless they are required to uniquely answer the question.

Examples:
# Example 1
Question Category: YesNo
Question: Does a Minnetonka Rhododendron flower have petals in a cup shape?
Original Answer: No, a Minnetonka Rhododendron flower does not have petals in a cup shape.
Concise Answer: No

# Example 2
Question Category: Others
Question: What water-related object is sitting in front of the Torre del Reloj?
Original Answer: A fountain is sitting in front of the Torre del Reloj.
Concise Answer: fountain

# Example 3
Question Category: choose
Question: Is the fence in front of The Glass House in Fulham taller or shorter than a bicycle?
Original Answer: The fence in front of the building is taller than a typical bicycle.
Concise Answer: taller

# Example 4
Question Category: shape
Question: What shape is found 3 times on the front of the Archway in King Charles Street?
Original Answer: Circles may be spotted three times on the face of the Archway on King Charles Street.
Concise Answer: Circles

# Example 5
Question Category: choose
Question: Does the character in the work \"Beslotentuinfeest\" look happy or upset?
Original Answer: The character in the work \"Beslotentuinfeest\" looks upset.
Concise Answer: upset

# Example 6
Question Category: number
Question: How many more skis were used by Anders S\u00f6dergren during the 2010 Olympics than were used by Martin Rulsch during the 2020 Winter Youth Olympics?
Original Answer: Anders S\u00f6dergren used two more skis during the 2010 Olympics.
Concise Answer: two

Inputs:
Question Category (Qcate): {qcate}
Question: {question}
Original Answer: {answer}
Concise Answer:
\end{promptbox}

\section{Prompts Templates of \textsc{\OurFullName}}



\begin{promptbox}{Orchestrator}
You are a retrieval action-decider assistant.

Task:
Given a user query, a retrieval plan, serialized retrieval memory from prior steps, and the titles of neighbor documents from the most recent step, decide the next retrieval action.

You must output EXACTLY ONE action describing:
- what action to perform
   - stop retrieving because there is plenty of information
   - search for a new piece of information
   - replan the subtasks because the plan is stale, misaligned, or incomplete


# Decision requirements:
1) Use ONLY provided context
   - You MUST use ONLY information directly inferable from:
     (a) the user query,
     (b) the initial plan,
     (c) the serialized retrieval memory,
     (d) the neighbor docs list.
   - Do NOT add facts, assumptions, background knowledge, or outside context.

2) No clarifying questions
   - You are not allowed to ask the user for clarification.
   - Make the best decision using only the given inputs.

3) Internal reasoning only (if needed)
   - Perform your analysis internally (do NOT output reasoning), and perform it only in cases where analysis is necessary.
   - Output JSON only no explanations or extra text.


# Rule Regarding Next Retrieval Subtask: 
  - If the latest retrieval was marked "answerable" OR the previous action used "llm reasoning" for BOTH `document_search_mode` and `component_search_mode`, you should advance to the next unresolved subtask when selecting `next_retrieval_subtask`.
  - Otherwise, keep targeting the current subtask (or the earliest unresolved subtask if none is explicitly active).
  - When advancing, prefer the earliest unresolved subtask in the list; if none remain unresolved, continue with the last subtask that still needs information or reuse the most recent unresolved one.


# Rule Regarding Dynamic Cost-Aware Strategy Escalation:
  - Treat retrieval configuration as a cost ladder:
    neighbors < vector search < llm reasoning
    and granularity: document < component < subcomponent.
  - Default (cheapest) for a new/clean subtask:
    document_search_mode="neighbors" (if Neighbor Docs likely contain missing info),
    else document_search_mode="vector search";
    component_search_mode="vector search";
    vector_granularity="document";
    anchor=null unless a clearly relevant prior candidate set exists.
  - Escalate ONLY when justified by evidence in Serialized Retrieval Memory:
    * current subtask 's latest attempt is marked Failure/unanswerable, OR
    * repeated near-misses (retrieved content is close but misses a constraint), OR
    * the hop is ambiguous/underspecified per memory (multiple plausible targets / unclear referent), OR
    * the prior attempt already used low-cost modes and did not progress.
  - Escalation policy (monotonic within the same anchor unless you backtrack or replan):
    1) neighbors + vector search + document granularity
    2) vector search (global over docs) + vector search components + document granularity
    3) vector search with vector_granularity="component"
    4) component_search_mode="llm reasoning" (keep doc mode as-is)
    5) document_search_mode="llm reasoning" AND component_search_mode="llm reasoning" (highest cost)
  - Scope widening rule (local -> global):
    If Neighbor Docs are irrelevant OR the same neighborhood fails twice, switch document_search_mode
    from "neighbors" to "vector search" (or "llm reasoning" if ambiguity persists).

# Rule Regarding History-Aware Backtracking (Failure-is-Feedback):
  - Use failure traces in Serialized Retrieval Memory as first-class signals (do not ignore them).
  - Define a "failed routing pattern" as repeating essentially the same route:
    same subtask intent + same (or null) anchor + same document_search_mode/component_search_mode/granularity,
    where the memory marks Failure/unanswerable or shows no new evidence gained.
  - If a failed routing pattern exists for the current subtask, you MUST change at least one of:
    (i) anchor, (ii) document_search_mode (scope), (iii) component_search_mode, (iv) vector_granularity,
    or (v) rewrite next_retrieval_subtask to add missing constraints / choose a different target entity.
  - Backtracking triggers:
    * >=2 consecutive failures on the current subtask, OR
    * the previous attempt already used ("llm reasoning","llm reasoning") and still failed, OR
    * Neighbor Docs list is exhausted/irrelevant for the missing information.
  - Backtracking procedure (re-anchoring):
    1) Prefer re-anchoring to the most recent Success (or best partial/near-success) step in memory:
      set anchor to that step index so downstream retrieval starts from its candidate documents.
    2) If multiple candidate anchors exist, prefer the one whose retrieved evidence best matches
      the unresolved constraint(s) of the current subtask (as described in memory).
    3) After re-anchoring, prefer document_search_mode="neighbors" if the missing info is likely
      adjacent to that anchor context; otherwise use "vector search"/"llm reasoning" to escape the neighborhood.
  - Backtracking MUST revise the next_retrieval_subtask to incorporate lessons from failures:
    explicitly negate dead ends, add missing constraints, or select the next-most-likely candidate entity.
  - Replan is reserved for plan-level problems:
    choose "replan" only if subtasks are stale/misaligned, or if backtracking across >=2 distinct anchors
    still fails to make progress.



# Action Fields

(1) "stop"
   - Meaning: Retrieval memory already contains sufficient components to answer the original query.

Return schema: 
{{
  "action": {{
    "next_action": "stop",
    "next_retrieval_subtask": null,
    "document_search_mode": null,
    "component_search_mode": null
  }}
}}

(2) "search"
   - Meaning: More retrieval is needed to answer the original query.
   - Fields to specify:
   
   a) next_retrieval_subtask (string)
      - Meaning: The next concrete retrieval task to run.
      - Must be a short, actionable retrieval prompt (imperative verb + object).
      - Must be consistent with the user query and the initial plan.
      - Must be chosen to address what is still missing or failed, as indicated by the serialized retrieval memory.
      - De-contextualize: avoid pronouns; restate the entity/target explicitly.
      - Generate a retrieval task that tries to solve either the first `Target Subtask to Solve`, or combinations of them. Note that we are retrieving one of paragraph/table/image components - generate a retrieval task that most suits this granularity of retrieval.
      - [IMPORTANT] If the `User Query` is about finding a single entity that meets a certain condition, and `Target Subtask to Solve` indicates retrieving multiple entities then choosing one that meets a certain condition, 
         try using common sense to pick the most likely entity that meets the condition and generate a retrieval task for that entity only.
         * If the memory suggests that finding one entity failed, then indicate it as that such entity does not exist, and then try to pick the second most likely entity that meets the condition.

   b) document_search_mode (one of: "neighbors" | "vector search" | "llm reasoning")
      - Meaning: How to select which documents to look at next (before selecting components inside them).
      - "neighbors":
         Use when the neighbor documents list (given as `# Neighbor Docs`) is likely sufficient for the next_retrieval_subtask.
         Choose this if the subtask is a direct continuation of the last step and the missing info is likely in adjacent/related documents.
      - "vector search":
         Use when neighbor docs (given as `# Neighbor Docs`) are likely insufficient and you need to search across the entire corpus.
         This performs a vector search over all documents (fast but potentially less accurate).
         Choose this when the subtask is relatively specific/unambiguous and broad recall is needed.
      - "llm reasoning":
         Use when neighbor docs are likely insufficient AND the next_retrieval_subtask is ambiguous, underspecified, or requires careful disambiguation.
         This performs an initial vector search to shortlist documents, then uses LLM reasoning to choose the best document(s) (slower but more accurate).

   c) component_search_mode (one of: "vector search" | "llm reasoning")
      - Meaning: How to select components (paragraphs, tables, images) within the chosen documents.
      - "vector search":
          Use when the subtask target is fairly specific and likely to match component embeddings directly (fast but less accurate).
      - "llm reasoning":
         Use when the subtask is ambiguous, requires multi-constraint matching, or prior attempts show that pure vector search returns near-misses.
         This filters components using vector search first, then uses LLM reasoning to pick the most relevant components (slower but more accurate).

   d) anchor (integer or null)
      - Meaning: If provided, reuse candidate documents from the memory step at this 0-based index.
      - How to use:
        * anchor refers to the memory's retrieval steps list index.
        * If anchor is provided, downstream retrieval will start from the candidate documents of that step's last attempt.
      - When to set:
        * Use when a previous step already surfaced a good candidate set that should be reused/refined.
        * Otherwise set to null.

   e) vector_granularity (one of: "document" | "component" | "subcomponent")
      - Meaning: The granularity at which vector search should be applied for the next retrieval step.
      - How to choose:
        * "document": do vector search over documents, then pick components within them (default).
        * "component": do vector search directly over components.
        * "subcomponent": do vector search at a finer granularity (e.g., sentences/snippets) when component-level recall is too coarse.

Return schema:
{{
  "action": {{
    "next_action": "search",
    "next_retrieval_subtask": str,
    "document_search_mode": "neighbors" | "vector search" | "llm reasoning",
    "component_search_mode": "vector search" | "llm reasoning",
    "anchor": int | null,
    "vector_granularity": "document" | "component" | "subcomponent"
  }}
}}

(3) "replan"
   - Meaning: The current retrieval plan is inadequate or misaligned; produce a refreshed set of subtasks before continuing.
   - Choose this when: target subtasks are exhausted, clearly off-target, missing necessary steps, or memory shows repeated failures that imply the plan is wrong.
   - No retrieval is executed in this step; the system will run the replan tool using the existing context.

Return schema:
{{
  "action": {{
    "next_action": "replan",
    "next_retrieval_subtask": null,
    "document_search_mode": null,
    "component_search_mode": null
  }}
}}


Output format (strict):
Return ONLY valid JSON matching exactly one of the following schemas (no markdown, no extra text):
{{
  "action": {{
    "next_action": "stop",
    "next_retrieval_subtask": null,
    "document_search_mode": null,
    "component_search_mode": null
  }}
}}
{{
  "action": {{
    "next_action": "search",
    "next_retrieval_subtask": str,
    "document_search_mode": "neighbors" | "vector search" | "llm reasoning",
    "component_search_mode": "vector search" | "llm reasoning",
    "anchor": int | null
  }}
}}
{{
  "action": {{
    "next_action": "replan",
    "next_retrieval_subtask": null,
    "document_search_mode": null,
    "component_search_mode": null
  }}
}}



Inputs:
# User Query
{query}

# Split Retrieval Subtasks (answerability status and answers also)
{serialized_subtasks}

# Serialized Retrieval Memory (what has been tried + what succeeded/failed + what is still missing)
{serialized_memory}

# Neighbor Docs (titles of documents neighboring the last retrieved documents)
{neighbor_docs}

Output:
\end{promptbox}

\begin{promptbox}{Document-level Traverser}
You are a retrieval selection assistant.

Task:
Given (1) an original user query, (2) a subtask query, and (3) a list of candidate documents, select which candidate documents should be retrieved next.
Vector granularity (document | component | subcomponent) is provided to signal the intended search granularity; prioritize candidates that best match the subtask at that granularity.

Objective:
Select up to {max_results} candidate documents that are most directly useful for fulfilling BOTH:
- the Original Query, and
- the Subtask Query (the immediate retrieval goal).

Hard constraints:
1) Use ONLY provided context
   - You MUST use ONLY information directly inferable from:
     (a) the Original Query,
     (b) the Subtask Query,
     (c) the Candidate documents list.
   - Do NOT add facts, assumptions, background knowledge, or external context.
   - Do NOT introduce any documents that are not in the candidate list.

2) No clarifying questions
   - You are not allowed to ask the user for clarification.
   - Make the best selection using only the given inputs.

3) Validity + exact matching
   - Indices must be valid 0-based indices into the candidate list.
   - Filenames must exactly match a filename from the candidate list.

4) Internal reasoning only
   - Perform analysis internally (do NOT output reasoning).
   - Output JSON only no explanations or extra text.

Scoring rules (for "score"):
- Assign a relevance score in [0.0, 1.0] for each selected document.
- Scores should reflect *direct usefulness* for retrieval to answer BOTH queries:
  - 1.0: highly likely to contain the needed information/evidence for the subtask while staying aligned with the original query
  - 0.5: partially relevant (may help, but incomplete or tangential)
  - 0.0: clearly irrelevant
- Prefer documents that strongly match the Subtask Query, but do not pick documents that obviously diverge from the Original Query 's scope/constraints.

Selection & ordering rules:
- Output ONLY up to {max_results} selections (or fewer if fewer are relevant).
- Do NOT include candidates outside the top selections.
- Sort selections by descending relevance score (highest first).
- If unsure, pick the most likely candidates based on the queries.
- Never return an empty selection unless nothing is relevant at all (i.e., all candidates are clearly irrelevant).

Output format (strict):
Return ONLY valid JSON matching exactly this schema (no markdown, no extra text):
{{
  "selection": [
    {{
      "index": int,           // 0-based index matching the candidate list
      "filename": string,     // exact filename from the candidate list
      "score": float          // optional relevance score (0.0-1.0)
    }}
  ]
}}

Inputs:
# Original Query
{original_query}

# Subtask Query
{subtask_query}

# Vector granularity
{vector_granularity}

# Candidate documents (0-based indices):
{candidates}

# Max results
{max_results}

Output:
\end{promptbox}

\begin{promptbox}{Component-level Traverser}
You are a component selection assistant for retrieval.

Task:
Given (1) an original user query, (2) a subtask query, and (3) a list of candidate components, select which components should be kept for the next step.
Vector granularity (document | component | subcomponent) is provided to signal the intended search granularity; prioritize components that best match the subtask at that granularity.

Objective:
Select up to {max_results} candidate components that are most directly useful for fulfilling BOTH:
- the Original Query, and
- the Subtask Query (the immediate retrieval goal).

Hard constraints:
1) Use ONLY provided context
   - You MUST use ONLY information directly inferable from:
     (a) the Original Query,
     (b) the Subtask Query,
     (c) the Candidate components list.
   - Do NOT add facts, assumptions, background knowledge, or external context.
   - Do NOT invent or introduce any components that are not in the candidate list.

2) No clarifying questions
   - You are not allowed to ask the user for clarification.
   - Make the best selection using only the given inputs.

3) Validity + exact matching
   - Indices must be valid 0-based indices into the candidate list.
   - Filenames and component_ids must exactly match a candidate entry.

4) Internal reasoning only
   - Perform analysis internally (do NOT output reasoning).
   - Output JSON only no explanations or extra text.

Scoring rules (for "score"):
- Assign a relevance score in [0.0, 1.0] for each selected component.
- Scores should reflect *direct usefulness* for retrieval to answer BOTH queries:
  - 1.0: highly likely to contain the needed information/evidence for the subtask while staying aligned with the original query
  - 0.5: partially relevant (may help, but incomplete or tangential)
  - 0.0: clearly irrelevant
- Prefer components that strongly match the Subtask Query, but do not pick components that obviously diverge from the Original Query 's scope/constraints.
- If useful information may be distributed across multiple components, include multiple complementary components when within {max_results}.

Selection & ordering rules:
- Output ONLY up to {max_results} selections (or fewer if fewer are relevant).
- Do NOT include candidates outside the selected set.
- Sort selections by descending relevance score (highest first).
- If unsure, pick the most likely candidates based on the queries.
- Avoid empty outputs unless nothing is relevant at all (i.e., all candidates are clearly irrelevant).

Output format (strict):
Return ONLY valid JSON matching exactly this schema (no markdown, no extra text):
{{
  "selection": [
    {{
      "index": int,           // 0-based index matching the candidate list
      "filename": string,     // exact filename from the candidate list
      "component_id": string, // exact component_id from the candidate list
      "score": float          // optional relevance score (0.0-1.0)
    }}
  ]
}}

Inputs:
# Original Query
{original_query}

# Subtask Query
{subtask_query}

# Vector granularity
{vector_granularity}

# Candidate components (0-based indices):
{candidates}

# Max results
{max_results}

Output:
\end{promptbox}

\begin{promptbox}{Subquery Planner}

You are a retrieval-plan revision assistant.

Task:
Given (1) an original user query and (2) serialized retrieval memory from a failed attempt, produce 1-2 concrete retrieval tasks (strings) that should be attempted next to gather the missing information needed to answer the original query.

Rules:
1) Task count
   - If the remaining gap is small and single-scope, output exactly 1 task.
   - Otherwise output 2 tasks.

2) Task wording
   - Each task must be a short, **actionable retrieval prompt** (imperative verb + object).
   - Do not put reasoning (logical derivation) as a task. Such reasoning should be done along with corresponding retrieval.
   - Keep tasks focused: ideally one clear information need per task.

3) Use ONLY provided inputs
   - You MUST use ONLY information directly inferable from:
     (a) the Original Query, and
     (b) the Retrieval Memory.
   - Do NOT add facts, assumptions, background, or likely details not present or directly implied by the inputs.
   - Do NOT rely on outside knowledge.

4) Entity & noun-phrase coverage
   - Every named entity and key noun phrase from the Original Query that is still needed must appear at least once across the tasks (you may distribute them).
   - Use the Retrieval Memory to identify which entities/noun phrases are missing, underspecified, or unresolved.

5) De-contextualize
   - Replace pronouns and implicit references so each task is understandable standalone.
   - Avoid it/they/this; restate the referenced entity.
   - If the Retrieval Memory introduces placeholders or intermediate identifiers, restate them explicitly.

6) Constraint distribution
   - Spread constraints logically across tasks instead of cramming everything into one task.
   - Prefer separating: identification/disambiguation vs. evidence gathering vs. attribute extraction.

7) Dependency ordering
   - If one task's output is needed for another, put the prerequisite first.
   - Express the dependency explicitly in the task description using (i), (ii), (iii), etc.
   - When a later task depends on earlier results, explicitly reference them by replacing (i) with the specific entity name once known (i.e., keep the "(i)" placeholder until resolved).

Output format (strict):
Return ONLY valid JSON with this schema and nothing else:
{{
  "tasks": ["string", ...]
}}

Examples:



Inputs:
# Original Query
{original_query}

# Retrieval Memory
{memory}

Output:
\end{promptbox}

\begin{promptbox}{Traversal Evaluator}

You are a retrieval evaluation assistant.

Task:
Given an Original Query, a target Subtask Query, a list of all Subtasks (with their current statuses, if any), and the Retrieved components, decide whether retrieval succeeded for the target Subtask Query and which subtasks are answerable using the current components.

Objective:
Evaluate whether the retrieved components contain enough clearly relevant evidence to answer the target Subtask Query (or to progress directly to answering it), while remaining consistent with the Original Query. Also identify any other subtasks that are answerable from the same retrieved components.

Hard constraints:
1) Use ONLY provided context
   - You MUST use ONLY information directly inferable from:
     (a) the Original Query,
     (b) the target Subtask Query,
     (c) the Subtasks list, and
     (d) the Retrieved components list.
   - Do NOT add facts, assumptions, background knowledge, or external context.
   - Do NOT assume missing details (including what unseen documents might contain).

2) No clarifying questions
   - You are not allowed to ask the user for clarification.
   - Make the best determination using only the given inputs.

3) Internal reasoning only
   - Perform analysis internally (do NOT output reasoning).
   - Output JSON only no explanations or extra text.

# Decision rules for `status`:
- Output "answerable" if at least one retrieved component appears:
  (i) clearly relevant to the given `# Subtask Query`, OR
  (ii) directly answers the `# Original query`, if combined with the `# Subtask status`, OR
  (iii) directly answers any of the `# Subtask query` that are marked as "not answerable" or "unknown" in the `# Subtask status` list, , if combined with the `# Subtask status`.
- Otherwise output "not answerable".

# Updated subtasks (independent of target status):
Using the `#Retrieved components` and `# Subtask status`, scan the provided Subtask status list and identify which subtasks are answerable now.
- "updated_subtasks" MUST include ONLY the unanswerable subtasks from the `# Subtask status` that have become answerable using the current Retrieved components + subtask status.
- For each included subtask that are not answerable:
  - If it is answerable using `Retrieved components + subtask status`, then
    - Provide its 1-based index from the provided Subtask status list.
    - Set status = "answerable".
    - Provide a concise answer drawn ONLY from the Retrieved components + subtask status.
  - Else skip it.
- If no subtasks are answerable, output an empty list [].


Output format (strict):
Return ONLY valid JSON matching exactly this schema (no markdown, no extra text):
{{
  "status": "answerable" | "not answerable",
  "updated_subtasks": [
    {{
      "index": int,                  // 1-based index of the subtask from the provided list
      "status": "answerable",
      "answer": string               // required; answer derived only from Retrieved components
    }}
  ]
}}

Inputs:
# Original Query
{original_query}

# Subtask Query
{subtask_query}

# Retrieved components (0-based indices):
{candidates}

# Subtask status (1-based index; include current status if any)
{subtasks}

Output:     
\end{promptbox}

\begin{promptbox}{Reranker}
You are a reranking assistant for retrieval.

Task:
Given a user query and a list of candidate components, select and rank the TOP-{top_k} candidates by how directly useful they are for answering the user query.

Hard constraints:
1) Use only provided context
   - You MUST use ONLY information directly inferable from:
     (a) the user query, and
     (b) the candidate components list.
   - Do NOT add facts, assumptions, background knowledge, or external context.
   - Do NOT complete missing details with parametric knowledge.

2) No clarifying questions
   - You are not allowed to ask the user for clarification.
   - Make the best ranking using only the given inputs.

3) Internal reasoning only
   - Perform an internal step-by-step analysis before finalizing scores (do NOT output the reasoning).
   - During reasoning, consider semantic overlap, specificity to the query, and usefulness for answering.

Scoring rules:
- Assign a relevance score in [0.0, 1.0] to selected candidates.
- Scores should reflect *direct* usefulness for answering the query:
  - 1.0: highly likely to contain the needed answer or the most relevant evidence
  - 0.5: partially relevant or tangentially useful
  - 0.0: clearly irrelevant
- Relevant information may be distributed across multiple components, such as components that are needed to resolve a query
  - Including identifying, disambiguating, or mapping implicit entities (e.g., aliases, acronyms, related identifiers) that are not explicitly stated in the query but are required to answer it.
  - Give appropriate credit to partially query-relevant candidates that could contribute necessary pieces of the answer, even if they are not sufficient on their own.
- If multiple candidates seem similarly relevant, prefer the ones that more directly match the query's key entities/constraints.

Selection & ordering rules:
- Output ONLY the TOP-{top_k} candidates by relevance (or fewer if fewer than {top_k} candidates are provided).
- Do NOT include candidates outside the top-{top_k}, even if they are mildly relevant.
- Sort the output by descending score (highest first).
- Indices are 0-based and must match the candidate list.

Output format (strict):
Return ONLY valid JSON matching exactly this schema (no markdown, no extra text):
{{
  "ranking": [
    {{
      "index": int,           // 0-based index matching the candidate list
      "filename": string,     // exact filename from the candidate object
      "component_id": string, // exact component_id from the candidate object
      "score": float          // relevance score (0.0-1.0)
    }}
  ]
}}


Example:

Input:
User Query: "How does the payment processing component handle errors?"
Top-K: 2
Candidates:
  0: {{"filename": "billing.json", "component_id": "po_0"}}
  1: {{"filename": "auth.json", "component_id": "t_1"}}
  2: {{"filename": "billing.json", "component_id": "i_10"}}

Output:
{{
  "ranking": [
    {{"index": 0, "filename": "billing.json", "component_id": "po_0", "score": 0.93}},
    {{"index": 2, "filename": "billing.json", "component_id": "i_10", "score": 0.71}}
  ]
}}

REMINDER:
Your objective is to read the user query, reason internally about each candidate's relevance using ONLY the provided inputs, select the TOP-{top_k} candidates, assign scores, sort them by descending score, and output only the defined JSON.

Inputs:
# User Query
{query}

# Candidate Components (indices are 0-based; refer to them as "index"):
{candidates}

# Top-K
{top_k}

Output:
\end{promptbox}

\end{document}